\documentclass[reprint,aps,pre,showpacs,floatfix,amsmath]{revtex4-1}
\usepackage{graphicx}
\usepackage{epsfig}
\usepackage{subfig}
\usepackage{paralist}

\newcommand{\eref}[1]{(\ref{#1})}
\newcommand{\Fref}[1]{Figure \ref{#1}}

\newcommand{\rme}{\mathrm{e}}

\begin{document}

\title{Numerical simulations of the Ising model on the Union Jack lattice}

\author{Vincent Mellor}
\email{vincent.mellor@uqconnect.edu.au}

\author{Katrina Hibberd}
\email{keh@maths.uq.edu.au}

\affiliation{Centre of Mathematical Physics, The University of Queensland, Queensland, 4072, Australia}

\begin{abstract}
This paper reviews the work of Wu and Lin on the Union Jack lattice Ising model. This model is of interest as it one of the few to display re-entrant phase transitions. Specifically we re-examine their result for the general anisotropic sublattice magnetisations, comparing these with the works of  Vaks, Larkin and Ovchinnikov, and our own numerical simulations. We discuss the disagreements found in both sublattice predictions including non-zero antiferromagnetic results and a rotational variance. We will then suggest additional conditions and modified formulae that will allow valid results to be produced.
\end{abstract}



\maketitle

\section{Introduction}
\label{sec:Intro}

\begin{figure}[htbp]
	\centering
		\includegraphics[width=0.75\columnwidth]{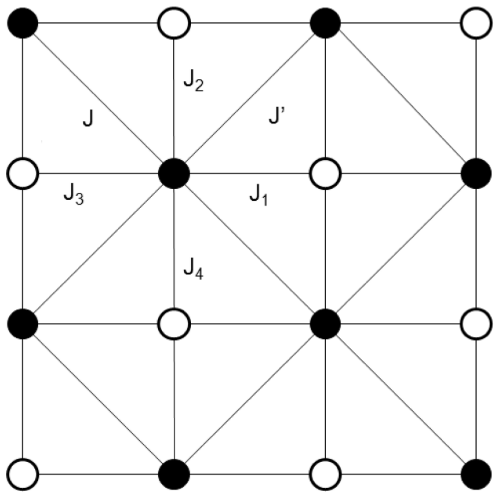}
	\caption{The Union Jack Lattice. (From \cite{Wu1987})}
	\label{fig:union}
\end{figure}

The Ising model is the prototypical model of phase transitions and as such is greatly studied \cite{Plischke1994}. Is is made up of sites interconnected along a lattice of ``bonds". Each of these sites can have spin of value $\pm 1$. The Union Jack lattice is obtained by adding alternate diagonals to a square lattice as shown in \Fref{fig:union}. This lattice is made up of two sublattices, the $\sigma$-lattice with sites with eight intersite ``bonds'' (filled circles) and the $\tau$-sublattice with sites with four of these ``bonds'' (hollow circles).

The Union Jack lattice Ising model is of particular interest as it is one of the few exactly solvable models which exhibits a re-entrant phase transition \cite{Vaks1966}. A solution was presented for the general anisotropic model on this lattice for the $\sigma$-sublattice in \cite{Wu1987} and then later for the complete lattice in \cite{Wu1989}. The solution showed that the Union Jack lattice Ising model is equivalent to the free fermion model. The spontaneous magnetisation for the free fermion model had been computed in \cite{Baxter1986}. The Union Jack lattice Ising model has been greatly studied since these solutions, with the work being extended in \cite{Strecka2006,Strecka2006a} to study a mixed spin lattice.

This paper focuses on the numerical simulations of the work in the thesis on the general anisotropic Union Jack Ising model in \cite{Mellor2010}. To start, in Section \ref{sec:UJL} we will state the required background for the paper. In this section we will also present the sublattice prediction functions from the work of \cite{Wu1987,Wu1989}. In Section \ref{sec:Res} we will briefly discuss the method used in our simulation program, along with the qualitative modelling of the Union Jack Ising model. We will then go onto present a theoretical analysis of the work presented in \cite{Wu1987,Wu1989} and \cite{Vaks1966} and compare these results with those from our numerical simulations. We will show that the prediction functions of \cite{Wu1987,Wu1989} do not accurately model the results of many of the anisotropic systems with invalid results being produced. Specifically we will identify the cases where the $\tau$-sublattice predictions are physically implausible and those which are not consistent under rotation of the lattice. In the conclusion in Section \ref{sec:con} we will state the additional conditions required to allow the predictions to produce valid results and the causes of the rotational variance.

\section{The Union Jack Lattice}
\label{sec:UJL}
In this section we will briefly state the equations for spontaneous magnetisation for the general anisotropic Ising model on the Union Jack lattice as presented by Wu and Lin \cite{Wu1987,Wu1989}. For convenience we adopt the notation of Wu and Lin \cite{Wu1987} and label the nearest interaction strengths as $-J_r$, which are defined to be one of six values: $-J_1$, $-J_2$, $-J_3$, $-J_4$, $-J$, $-J'$. The resultant Boltzmann factors are given by
\begin{eqnarray}
	\omega \left( a,b,c,d\right) &=&2\exp\left[ \frac{\beta J\left( ab+cd\right) }{2}+\frac{\beta J^\prime \left(ad+bc\right) }{2}\right] \nonumber\\ && \times\cosh \left( a\beta J_1+b\beta J_2+c\beta J_3+d\beta J_4\right),
	\label{eq:ino22}
\end{eqnarray} 
where $\beta=1/k_BT$, $k_B$ is the Boltzmann constant and $a$,$b$,$c$ and $d$ are the four sites surrounding a $\tau$ site. This equation produces sixteen possible factors, which can be reduced by symmetry to eight distinct expressions \cite{Wu1987}:
\begin{eqnarray*}
	\omega _1=\omega \left( ++++\right)&=&2\rme^{\beta J+\beta J^\prime } \cosh \left( \beta \left( J_1+J_2+J_3+J_4\right)\right) \nonumber \\
	\omega _2=\omega \left( +-+-\right) &=&2\rme^{-\beta J-\beta J^\prime }\cosh \left( \beta \left( J_1-J_2+J_3-J_4\right) \right) \nonumber \\
	\omega _3=\omega \left( +--+\right) &=&2\rme^{-\beta J+\beta J^\prime }\cosh \left( \beta \left( J_1-J_2-J_3+J_4\right) \right) \nonumber \\
	\omega _4=\omega \left( ++--\right) &=&2\rme^{\beta J-\beta J^\prime }\cosh \left( \beta \left( J_1+J_2-J_3-J_4\right) \right) \nonumber \\
	\omega _5=\omega \left( +-++\right) &=&2\cosh \left( \beta \left( J_1-J_2+J_3+J_4\right) \right) \nonumber \\
	\omega _6=\omega \left( +++-\right) &=&2\cosh \left( \beta \left( J_1+J_2+J_3-J_4\right) \right) \nonumber \\
	\omega _7=\omega \left( ++-+\right) &=&2\cosh \left( \beta \left( J_1+J_2-J_3+J_4\right) \right) \nonumber \\
	\omega _8=\omega \left( -+++\right) &=&2\cosh \left( \beta \left( -J_1+J_2+J_3+J_4\right) \right).
	\label{eq:ino23}
\end{eqnarray*}
In \cite{Wu1987} it is shown that the Union Jack lattice is equivalent to an eight-lattice models with weights given by \eref{eq:ino22}. This eight lattice model satisfies the free fermion condition \cite{Fan1970}. The spontaneous magnetisation of a free fermion model was given by Baxter in \cite{Baxter1986}. As such, the spontaneous magnetisation for the $\sigma$-sublattice is
\begin{eqnarray}
	\left\langle \sigma \right\rangle &=&\left\{
	\begin{array}{cc} 
		\left(1-\Omega^{-2}\right)^{1/8}, & \Omega^{-2}\geq 1 \\
		0, & \Omega^{-2} \leq 1,
	\end{array}
	\right.
	\label{eq:ino25} \\ \nonumber \\
	\Omega^2 &=&1-\frac{\gamma_1 ~ \gamma_2 ~  \gamma_3 ~ \gamma_4}{16\omega_5 ~\omega_6 ~\omega_7 ~\omega_8},
	\label{eq:uno3}
\end{eqnarray}
where
\begin{eqnarray*}
	\gamma_1 &=& -\omega_1+\omega_2+\omega_3+\omega_4 \nonumber \\
	\gamma_2 &=& \omega_1-\omega_2+\omega_3+\omega_4 \nonumber \\
	\gamma_3 &=& \omega_1+\omega_2-\omega_3+\omega_4 \nonumber \\
	\gamma_4 &=& \omega_1+\omega_2+\omega_3-\omega_4.
	\label{eq:uno1}
\end{eqnarray*}
The critical point(s) of this system are
\begin{displaymath}
	\Omega^2=1
\end{displaymath}
or equivalently,
\begin{equation}
	\omega_1+\omega_2+\omega_3+\omega_4=2\ \mathrm{max}\left\{\omega_1,\omega_2,\omega_3,\omega_4\right\}.
	\label{eq:uno5}
\end{equation}

The $\tau$-sublattice magnetisation is given by \cite{Wu1989} as
\begin{equation}
	\left\langle \tau \right\rangle = \left\langle \sigma \right\rangle \left[A_{1234}(K)(F_+ + F_-)+A_{2341}(K)(F_+ - F_-)\right].
	\label{eq:ino68}
\end{equation}
We can see that this equation is a multiple of the $\sigma$-sublattice value. In (\ref{eq:ino68}) from \cite{Wu1989},
\begin{eqnarray}
	A_{1234}(K)&=& \frac{\sinh{2(\beta J_1+\beta J_3)}}{\sqrt{2G_{-}(\beta J)\sinh{2\beta J_1}\sinh{2\beta J_3}}} \nonumber \\
	A_{2341}(K)&=& \frac{\sinh{2(\beta J_2+\beta J_4)}}{\sqrt{2G_{-}(\beta J)\sinh{2\beta J_2}\sinh{2\beta J_4}}} \label{eq:change}
\end{eqnarray}
and
\begin{displaymath}
	G_{-}(\beta J)= \cosh{2(\beta J_1+\beta J_3)}+\cosh{2(\beta J_2-\beta J_4)}.
\end{displaymath}
The calculation for $F_+$ and $F_-$ is a little more involved. We start by calculating
\begin{displaymath} 
	F_{\pm}= \sqrt{\frac{A+2\sqrt{BC}}{D+2E\sqrt{B}}}
\end{displaymath}
where
\begin{eqnarray*}
	A&=& 2\omega_5\omega_6\omega_7\omega_8\left(\omega^2_1+\omega^2_2+\omega^2_3+\omega^2_4\right) \nonumber \\
	&&-\left(\omega_1\omega_2+\omega_3\omega_4\right)\left(\omega_1\omega_3 + \omega_2\omega_4\right)\left(\omega_1\omega_4+\omega_2\omega_3\right) \nonumber \\
	B&=& \omega_5\omega_6\omega_7\omega_8\left(\omega_5\omega_6\omega_7\omega_8-\omega_1\omega_2\omega_3\omega_4\right) \nonumber \\
	C&=& \left(\omega_1^2+\omega_2^2+\omega_3^2+\omega_4^2\right)^2-4\left(\omega_5\omega_6-\omega_7\omega_8\right)^2 \nonumber \\
	D&=& \left(\omega_1^2+\omega_2^2\right)\left(2\omega_5\omega_6\omega_7\omega_8- \omega_1\omega_2\omega_3\omega_4\right) \\ && - \omega_5\omega_6\omega_7\omega_8\left(\omega_3^2+\omega_4^2\right) \nonumber \\
	E&=& \omega_1^2-\omega_2^2. \nonumber
\end{eqnarray*}
We can relate $F_+$ and $F_-$ with the following formula from \cite{Wu1989}, allowing us to get values for each variable,
\begin{displaymath}
	F_+F_- = \frac{\omega_5\omega_6-\omega_7\omega_8}{\omega_1\omega_2}.
\end{displaymath}
We can compute the overall nearest neighbour magnetisation by taking the mean of the two sublattice magnetisations
\begin{displaymath}
	M_0= \frac{1}{2}(\left\langle \sigma \right\rangle + \left\langle \tau \right\rangle).
\end{displaymath}

\subsection{Classification of phases}
\label{sub:CoPh}
In \cite{Wu1987}, a classification of the phase of the $\sigma$-sublattice is presented. It is based on the following energy values:
\begin{eqnarray*}
	-E_1 &=& J + J^\prime + \left|J_1+J_2+J_3+J_4\right| \nonumber\\
	-E_2 &=& -J - J^\prime + \left|J_1-J_2+J_3-J_4\right| \nonumber \\
	-E_3 &=& -J + J^\prime + \left|J_1-J_2-J_3+J_4\right|  \nonumber\\
	-E_4 &=& J - J^\prime + \left|J_1+J_2-J_3-J_4\right|.
	\label{eq:uno6}
\end{eqnarray*}
The sublattice is in:
\begin{asparaenum}[i)]
\item a ferromagnetic phase when $E_1 < E_2,~E_3,~E_4;$
\item a antiferromagnetic phase when $E_2 < E_1,~E_3,~E_4;$
\item a metamagnetic phase when $E_3 < E_1,~E_2,~E_4$ or $E_4 < E_1,~E_2,~E_3.$
\end{asparaenum}

As the temperature rises, depending on the relative strengths of the interactions $J_r$, the occurrence of a phase change is signified by one or more of the equations in \eref{eq:uno5} being realised. A re-entrant phase transition occurs if any one equation admits two solutions.

\section{Results}
\label{sec:Res}
In our work in \cite{Mellor2010} we presented a theoretical and numerical analysis of the results derived in \cite{Wu1987}. In our theoretical analysis we compared these results against those of \cite{Vaks1966} to identify systems to further investigate numerically. The results of both the theoretical and numerical analysis will be presented below. In \cite{Mellor2010} we also presented analysis of a mean-field approximation of the Union Jack lattice using two approaches. The first used a set partially uncoupled predictor equations, both functions of $\left\langle\sigma\right\rangle$. The second approximation used coupled predictor equations that are functions of $\left\langle\sigma\right\rangle$ and $\left\langle\tau\right\rangle$. Qualitatively we saw that our Mean Field models showed good correlation with the isotropic ferromagnetic systems. In anisotropic antiferromagnetic systems as well as anisotropic ferromagnetic systems with a re-entrant phase transition, the correlation was poor between the Mean Field results and the theoretical predictions.

As the theoretical results are for an infinite lattice, we used a Monte Carlo Markov Chain method with periodic boundary conditions to numerically simulate the systems. Our Monte Carlo algorithm was the Metropolis-Hastings algorithm \cite{Plischke1994, Metropolis1953}. For our simulations we will apply this algorithm to a lattice of 100 sites by 100 sites. This lattice size has been chosen as it is small enough to have a reasonable run time while being large enough to suppress the finite size effects. The computer program was calibrated against exact results for the general anisotropic triangular lattice for both the average magnetisation \cite{Stephenson1964} and the three-site correlator \cite{Baxter1975}. With this calibration we have confidence in the accuracy of the simulation results. The code for this simulation program can be found in \cite{Mellor2010}.

To classify the phase of the $\sigma$-sublattice we used a alternative approach to that given in Section \ref{sub:CoPh}. By using the individual $\gamma$ terms from \eref{eq:uno3} we note that the sublattice is in a ferromagnetic phase when $\gamma_1<0$, a antiferromagnetic phase when $\gamma_2<0$, a metamagnetic phase when $\gamma_3<0$ or $\gamma_4<0$ and a disordered phase when $\gamma_1\gamma_2\gamma_3\gamma_4>0$. The critical temperature(s) can also be determined when any of the $\gamma$ functions move from being negative to being positive, or vice versa.

\subsection{Isotropic ferromagnetic}
\label{sec:IF}
\begin{figure}[htbp]
	\centering
		\includegraphics[width= 0.9\columnwidth]{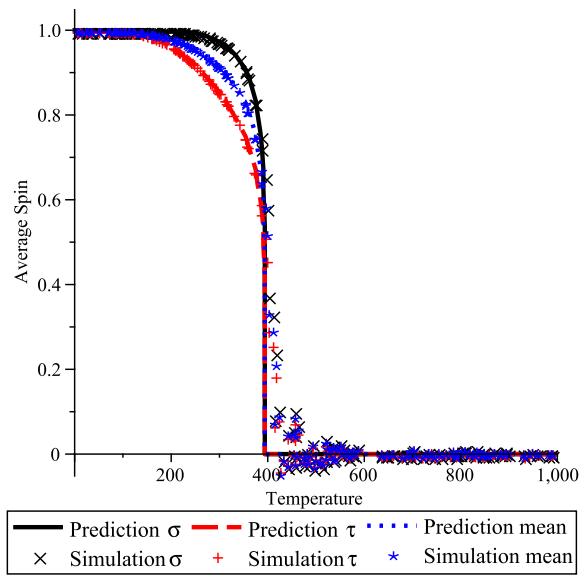}
	\caption{Plot of simulation results for an isotropic ferromagnetic system on the Union Jack lattice. Here the theoretical predictions are shown with the lines and the numerical results are shown with the points.}
	\label{fig:ferro}
\end{figure}

To start with we look at the isotropic ferromagnetic system. Here the interactions for our system will be $J=J^\prime=J_n=100k_B$. In our initial plot of this system the $\tau$-sublattice prediction \eref{eq:ino68} was a factor of two higher than expected. Upon analysis of the equation we found that due to the symmetric interactions of the system $A_{1234}=A_{2341}=1$, $F_+F_-=0$ and $\left\langle \tau \right\rangle = 2F_+\left\langle \sigma \right\rangle$. As such we adapted \eref{eq:ino68} to the following form
\begin{equation}
	\left\langle \tau \right\rangle = \left\langle \sigma \right\rangle \frac{\left[A_{1234}(K)(F_+ + F_-)+A_{2341}(K)(F_+ - F_-)\right]}{2}.
	\label{eq:papertau}
\end{equation}
As this is a minor adjustment we will continue to refer to the result as that of Wu and Lin \cite{Wu1989}. The plot of our numerical simulation results against the adapted prediction results of Wu and Lin \cite{Wu1987,Wu1989} are shown in Figure \ref{fig:ferro}.
Intuitively this is the type of graph we would expect for this type of system. As expected there is strong agreement between the two theories with the phase transition and critical temperature being the same. There is a high correlation between our simulation data and both sublattice prediction functions. There is some noise around the critical temperature, though it is of a small magnitude when compared to the other results. After the noise part of the data, around 400-600 Kelvin the simulation results again follow the prediction with a high correlation.

\subsection{Anisotropic metamagnetic}
\label{sec:NSmeta}
Next we move on to look at an anisotropic metamagnetic system where $J_n=10k_B$, $J=100k_B$ and $J^\prime=-100k_B$. The graph we obtain when we plot our simulation results against the predictions of Wu and Lin \cite{Wu1987, Wu1989} is shown in Figure \ref{fig:meta} below.
\begin{figure}[htbp]
	\centering
		\includegraphics[width= 0.9\columnwidth]{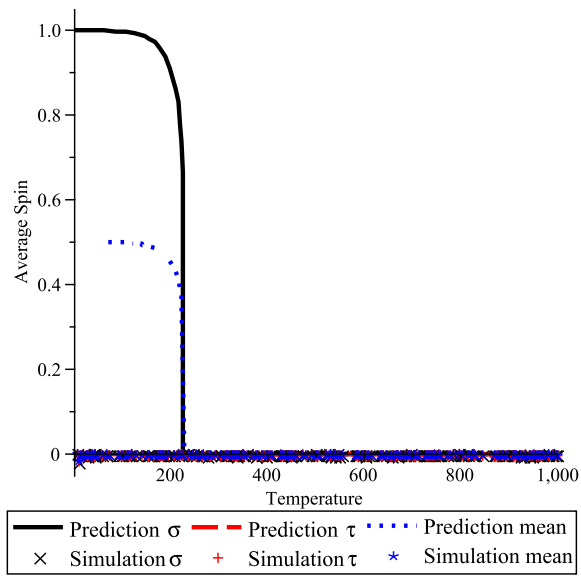}
	\caption{Plot of simulation results for an anisotropic metamagnetic system on the Union Jack lattice. Note that the simulation results do not follow the curves of the prediction functions.}
	\label{fig:meta}
\end{figure}
As can be seen from the graph, the prediction functions of \cite{Wu1987,Wu1989} show a non-zero magnetisation on the $\sigma$-sublattice. In this system the $\gamma_4$ is negative at low temperatures \cite{Mellor2010}. The equation for $\gamma_4$ is
\begin{displaymath}
 \gamma_4 = \omega_1+\omega_2+\omega_3-\omega_4,
\end{displaymath}
and we note from examination of this equation that $\omega_4=\omega \left( ++--\right)$ is the dominant term. From the spin configuration, this represents we would expect an antiferromagnetic phase with the average magnetisation for the $\sigma$-sublattice being zero. When we look at our simulation results for this sublattice we see that they have average spin zero which follows more closely the results we would expect.

\subsection{Anisotropic antiferromagnetic}
\label{sec:NSanti}
Next we study an anisotropic antiferromagnetic system. The chosen system of this type will have horizontal and vertical interactions of $J_n=100k_B$ and diagonal interactions of $J=J^\prime=-100k_B$. The simulation results are plotted against Wu and Lin's \cite{Wu1987,Wu1989} predictions in Figure \ref{fig:antiferro} below.
\begin{figure}[htbp]
	\centering
		\includegraphics[width= 0.9\columnwidth]{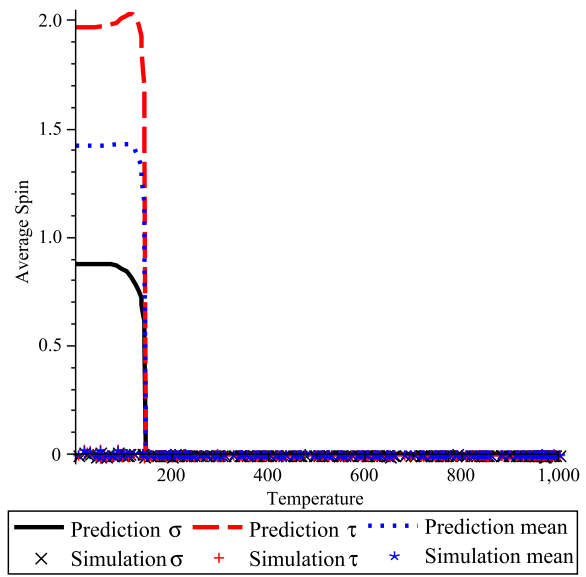}
	\caption{Plot of simulation results for an anisotropic antiferromagnetic system on the Union Jack lattice. Note that the simulation results do not follow the curves of the theoretical predictions which are physically impossible.}
	\label{fig:antiferro}
\end{figure}
Using similar analysis as in Section \ref{sec:NSmeta}, we find that $\omega_2=\omega(+-+-)$ is the dominant term for the magnetisation of the $\sigma$-sublattice. This configuration would agree with the result of \cite{Vaks1966} in that it predicts an ordered antiferromagnetic phase and the critical temperatures agree. However as we can see in \Fref{fig:antiferro}, the prediction function \eref{eq:uno3} produces a non-zero magnetisation for this phase where we would expect zero magnetisation.

When we examine the $\tau$-sublattice prediction function \eref{eq:papertau} we notice that it produces physically implausible results. That is, the predicted magnetisation of the $\tau$-sublattice is greater than the possible spin values, $\pm 1$, for this Ising model. As in the isotropic ferromagnetic case we have symmetric interactions leading to $A_{1234}=A_{2341}=1$ and $F_+F_-=0$. As in this case $F_+>2$, we can see that this is the cause of the implausible results.

The correlation between the simulation data and the predictions of \cite{Wu1987,Wu1989} is again poor. We see that our simulation results are centred around the zero magnetisation level, as we would intuitively expect. However it can be seen that the simulation results for the $\sigma$-sublattice do not show a difference between an ordered phase and a disordered phase.

\subsection{Anisotropic ferromagnetic}
\label{secNSanferro}
\begin{figure}[htbp]
	\centering
		\includegraphics[width= 0.9\columnwidth]{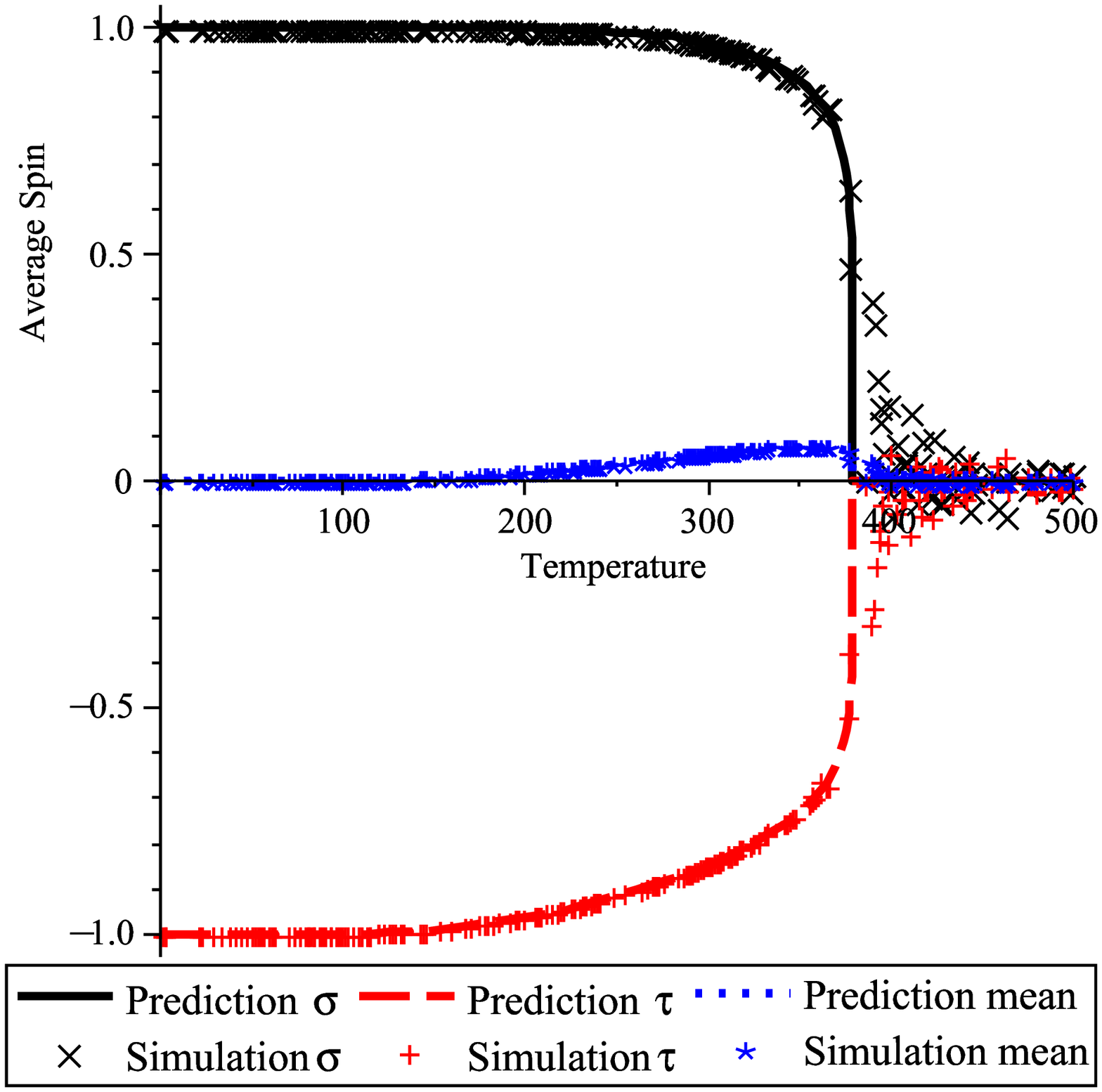}
	\caption{Plot of simulation results for an anisotropic ferromagnetic system on the Union Jack lattice with equal horizontal and vertical interactions.}
	\label{fig:antia92500}
\end{figure}

We now move to examining anisotropic systems with the $\sigma$-sublattice classified as ferromagnetic at low temperatures. This type of system is particularly interesting as it contains systems with staggered interactions and those with re-entrant phase transitions. We will also discuss rotational variance. First let us present a system containing staggered interactions, with the interactions of the system shown in \Fref{fig:antia92500} defined as $J_n=-100k_B$ and $J=J^\prime=92k_B$.
As we can see from the prediction functions of \cite{Wu1987,Wu1989} the average magnetisation have opposite sign. This agrees with the theoretical predictions for the critical temperature and state from \cite{Vaks1966}. It can also be noted that the overall magnetisation for the complete lattice has magnitude zero until around 200 Kelvin. The numerical simulation results have a high correlation with the theoretical predictions and follow the curves closely. There is noise after the critical temperature on both sublattices but this quickly reduces down to a small level for higher temperatures.

\begin{figure}[htbp]
	\centering
		\includegraphics[width= 0.9\columnwidth]{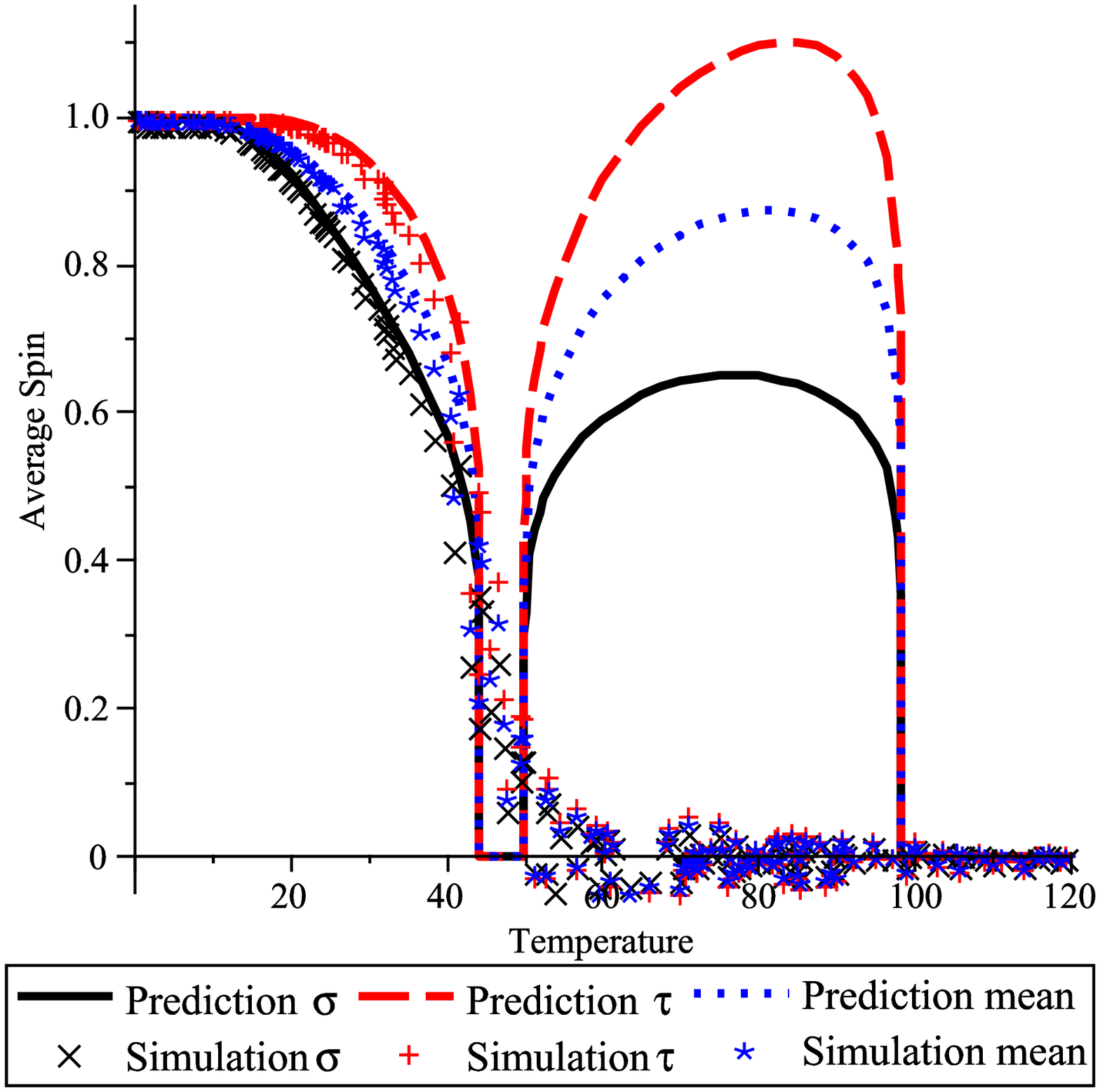}
	\caption{Plot of simulation results for an anisotropic ferromagnetic system on the Union Jack lattice with equal horizontal and vertical interactions.}
	\label{fig:anti92150}
\end{figure}

Having covered staggered interactions we now go onto looking at systems that contain a re-entrant phase transition. The system we studied was one with horizontal and vertical interactions of $J_n=100k_B$ and diagonal interactions of $J=J^\prime=-92k_B$. The graph of this system is shown in \Fref{fig:anti92150}.
This system is predicted by \cite{Vaks1966} to start in a ferromagnetic phase, move to a disordered phase, then into an ordered antiferromagnetic phase and then finally into a disordered phase again. As we see from the prediction functions of \cite{Wu1987,Wu1989}, initially the theories agree with a ferromagnetic phase being shown. The critical temperatures predicted by both theories agree in all three values. With further analysis we see that initially $\gamma_1$ is the dominant term, leading to a ferromagnetic phase. From the second critical temperature to the third we see that $\gamma_2$ is now the dominant term. However as we saw in Section \ref{sec:NSanti}, while we would expect a zero magnetisation, there is instead a non-zero value for both sublattices, with the $\tau$-sublattice having value greater than one.

In our numerical simulations we see that there is good correlation up to the second critical temperature but the simulation does not show the re-entrant phase transition. As we have discussed in Section \ref{sec:NSanti} both the ordered antiferromagnetic phase and disordered phase would have average magnetisation zero, and so are indistinguishable from this point of view.

Looking at more a general anisotropic system, we see that there exist systems which have rotational variance. As an example of these systems we first look at the system with horizontal interactions of $J_1=J_3=100k_B/0.9^2$, vertical interactions of $J_2=J_4=100k_B/0.9$ and diagonal interactions of $J=J^\prime=100k_B$, and  compare with the system with horizontal interactions of $J_1=J_3=100k_B/0.9$, vertical interactions of $J_2=J_4=100k_B/0.9^2$ (i.e. the lattice rotated through 90 degrees). The graphs of our simulation results are shown in Figure \ref{fig:NSfunkypos} below.
\begin{figure*}[htbp]
	\centering
		\subfloat[]{
		\includegraphics[width= 0.9\columnwidth]{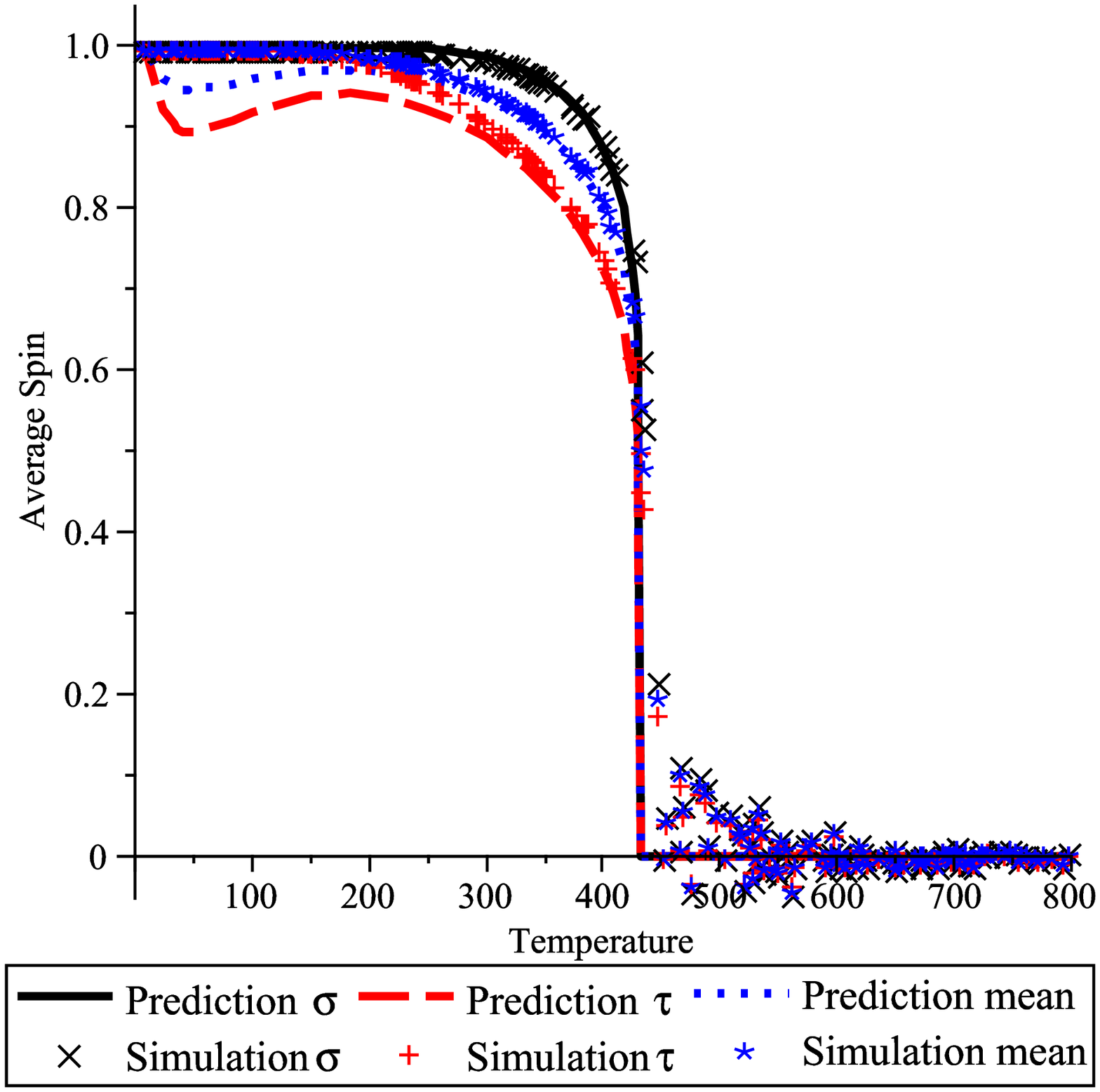}
		\label{subfig:funkygraphpos}
		}
		\subfloat[]{
		\includegraphics[width= 0.9\columnwidth]{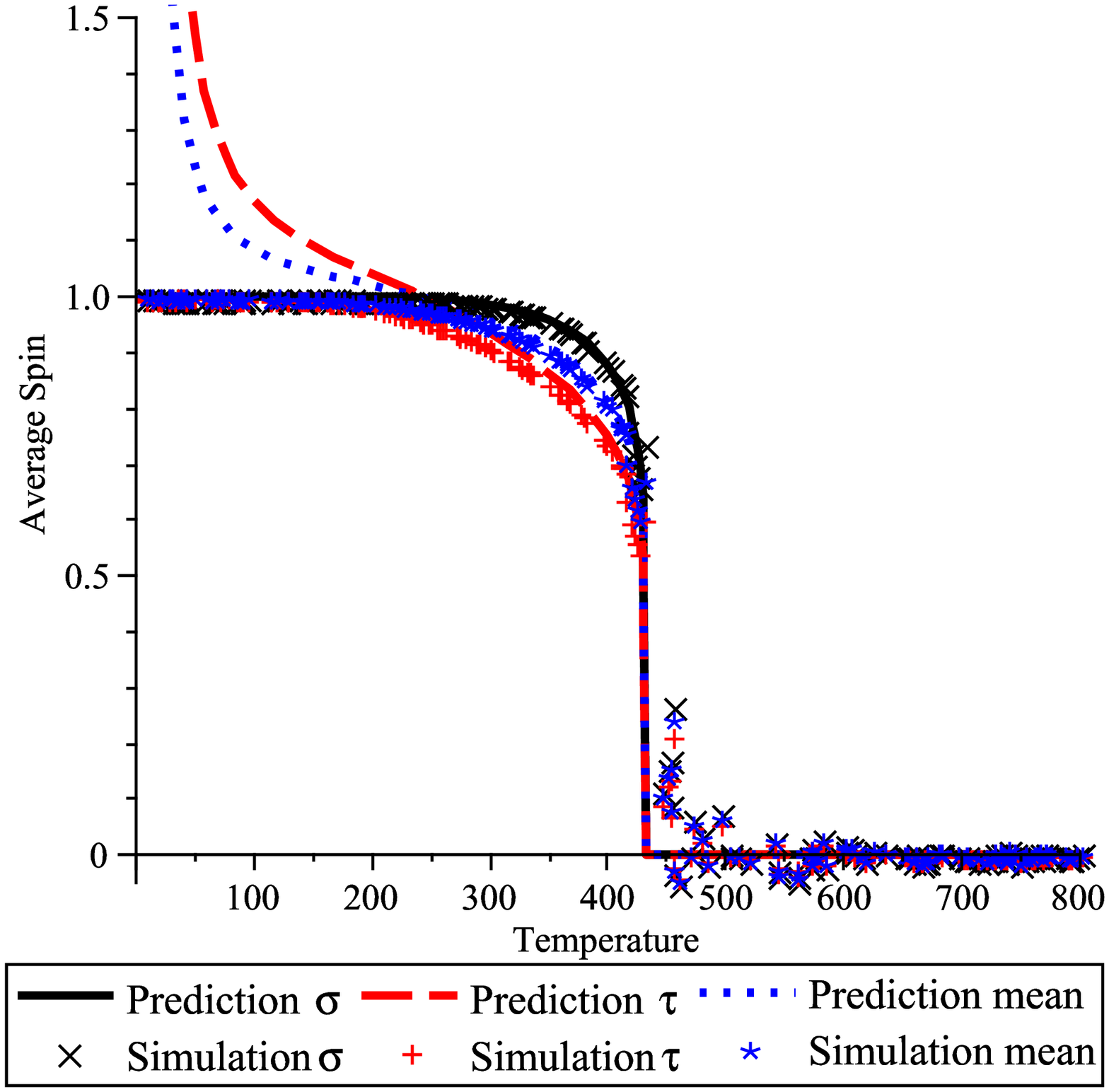}
		\label{subfig:funkyposgraph}
		}
	\caption{Plot of simulation results for an anisotropic ferromagnetic system on the Union Jack lattice with horizontal and vertical interactions that are not equal and positive diagonals. (a) shows the system with interactions $J_1=J_3=100k_B/0.9^2$, $J_2=J_4=100k_B/0.9$ $J=J^\prime=100k_B$. (b) shows the rotated system with interactions $J_1=J_3=100k_B/0.9$, $J_2=J_4=100k_B/0.9^2$ $J=J^\prime=100k_B$.}
	\label{fig:NSfunkypos}
\end{figure*}
As we see when we compare the simulations results, the data forms similar curves and has a similar phase transition at equal critical temperatures. In comparison to the predictions of Wu and Lin \cite{Wu1987,Wu1989}, we see that at higher temperatures the simulation results follow all three curves with good correlation. At lower temperatures, below about 200 Kelvin, we see that the $\sigma$ prediction still has good correlation for both systems. This suggests that the $\tau$-sublattice prediction is incorrect. Our simulation results show that rotation of the lattice should not have an effect on the results of the system. When we analyse \eref{eq:papertau} we see that
\begin{eqnarray}
F_+F_-&=& \frac{\cosh^2{2\beta J_{\mathrm{Horizontal}}}-\cosh^2{2\beta J_{\mathrm{Vertical}}}}{\cosh^2{2\beta J_{\mathrm{Horizontal}}}+\cosh^2{2\beta J_{\mathrm{Vertical}}}} \nonumber\\
A_{1234}&=& 1 \nonumber\\
A_{2341}&=& \frac{\cosh{2\beta J_{\mathrm{Vertical}}}}{\cosh{2\beta J_{\mathrm{Horizontal}}}}.
\label{eq:proof}
\end{eqnarray}
Both the rotational variance and disagreement between the $\tau$-sublattice and the simulation results are a result of \eref{eq:proof}.

\begin{figure*}[htbp]
	\centering
		\subfloat[]{
		\includegraphics[width= 0.9\columnwidth]{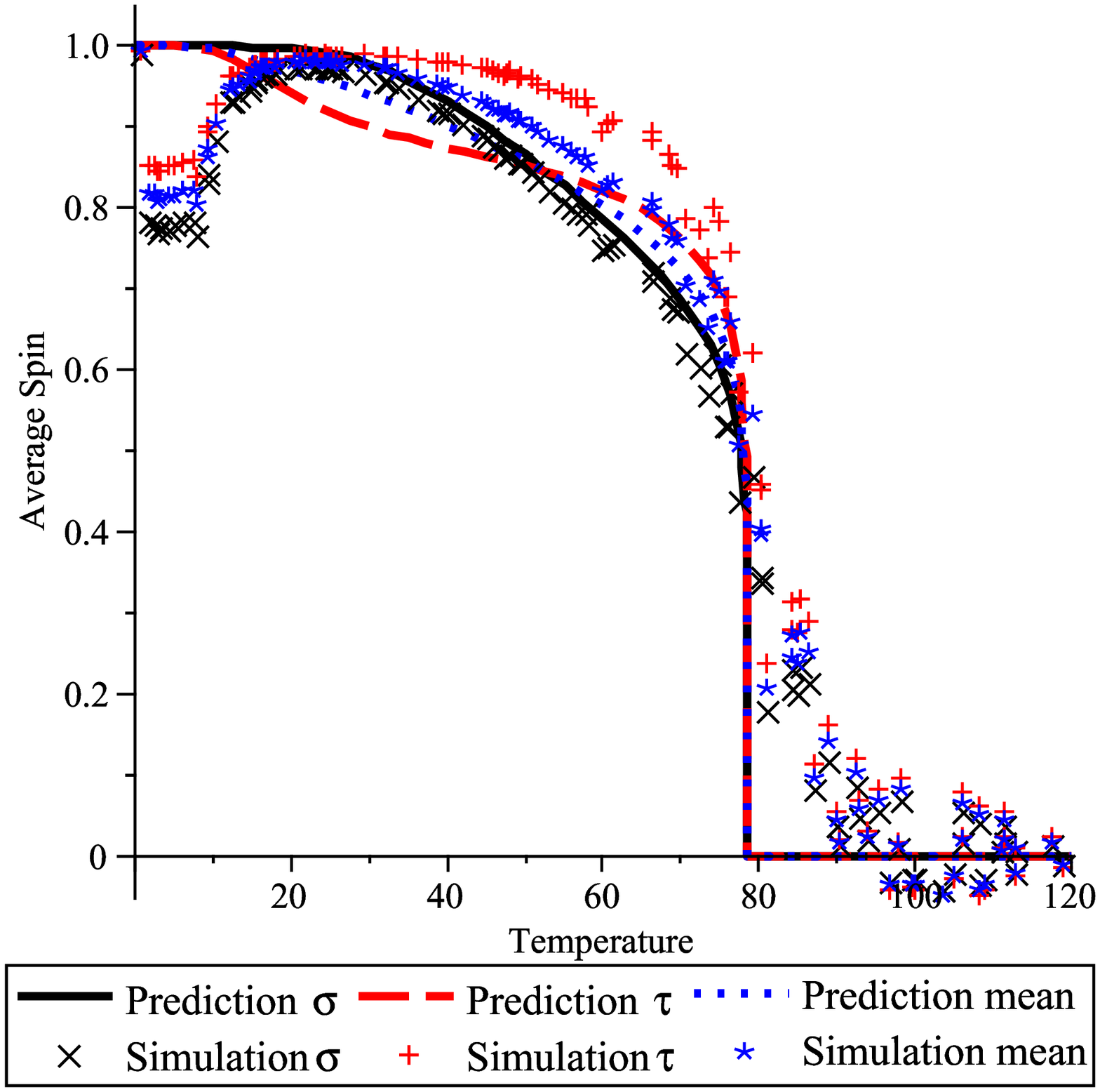}
		\label{subfig:funkydata1}
		}
		\subfloat[]{
		\includegraphics[width= 0.9\columnwidth]{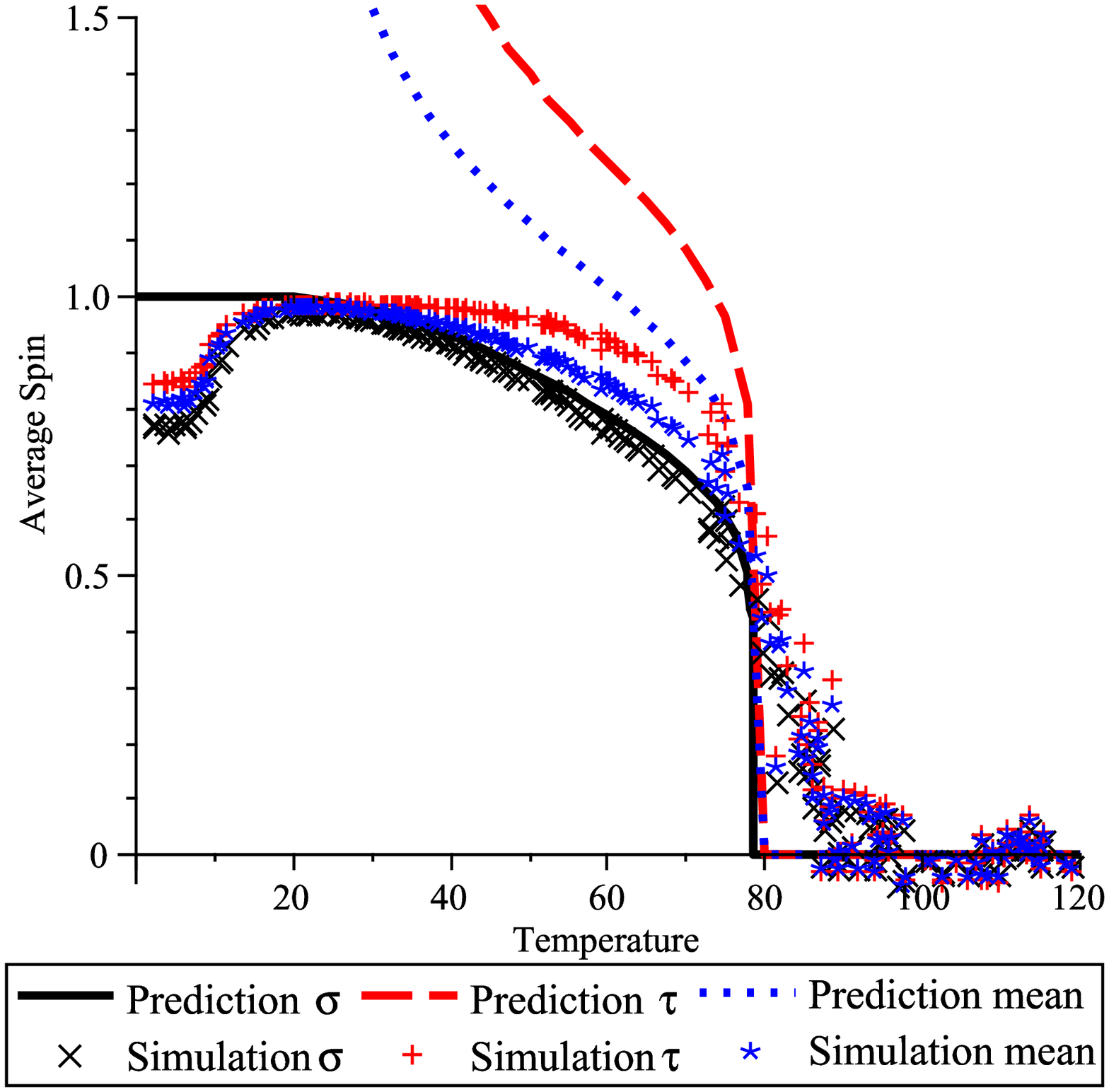}
		\label{subfig:funkydata2}
		}
	\caption{Plot of simulation results for an anisotropic ferromagnetic system on the Union Jack lattice with horizontal and vertical interactions that are not equal and negative values for the diagonal interactions. (a) shows the system with interactions $J_1=J_3=100k_B/0.9^2$, $J_2=J_4=100k_B/0.9$ $J=J^\prime=-100k_B$. (b) shows the rotated system with interactions $J_1=J_3=100k_B/0.9$, $J_2=J_4=100k_B/0.9^2$ $J=J^\prime=-100k_B$.}
	\label{fig:NSFunky}
\end{figure*}

A further result can be seen if we now take a system similar to the previous example but with negative diagonal interactions. For an example of this type of system, we will look at the system with horizontal interactions of $J_1=J_3=100k_B/0.9^2$, vertical interactions of $J_2=J_4=100k_B/0.9$ and diagonal interactions of $J=J^\prime=-100k_B$, and the same system rotated through 90 degrees. The graphs of our simulation results are shown in Figure \ref{fig:NSFunky}.
Again we see as in Figure \ref{fig:NSfunkypos}, that the simulation results are very similar to each other. However, in this case we see that below 10 Kelvin all three simulation results are lower than the predictions. After 10 Kelvin the simulation results again move up to the prediction curves, following the $\sigma$ prediction in both cases. We see that in both cases the simulation results and predicted results show the same critical temperature and phase transition. This oddity maybe due to the simulation being performed on a finite system, while the theoretical predictions are for an infinite lattice.

\section{Conclusion}
\label{sec:con}
We have seen that the re-entrant phase transitions of the Union Jack Ising model can not be seen when only considering the average magnetisation. This is due to the transition being from unordered to an ordered antiferromagnetic phase which both have an average magnetisation that is identically zero. In addition, the prediction of the $\sigma$-sublattice given in \cite{Wu1987} requires additional conditions to have agreement with our numerical simulations. It is possible to classify the phases of the system by examining the $\gamma_i$ terms of equation (\ref{eq:uno3}). The predictions with the current conditions produce non-zero magnetisations for non-ferromagnetic systems. However, if we impose the conditions that we only use the prediction formula (\ref{eq:ino25}) when $\gamma_1<0$ or $\gamma_1~\gamma_2~\gamma_3~\gamma_4~>0$ and it is zero outside those conditions, the results now work for all systems.

The prediction for the $\tau$-sublattice, given in \cite{Wu1989}, has additional issues in the general anisotropic lattice. Initially the prediction given in \cite{Wu1989} is too large by a factor of two, but can be easily corrected as shown in \eref{eq:papertau}. The $\sigma$-sublattice prediction is a factor of the $\tau$-sublattice prediction and so a zero magnetisation would be observed for the antiferromagnetic and metamagnetic phases if the above conditions are applied. The rotational variance seen in \Fref{fig:NSfunkypos} and \Fref{fig:NSFunky} can be eliminated by rewriting \eref{eq:change} as follows
\begin{eqnarray*}
	A_{1234}(K)&=& \frac{\sinh{2(\beta J_1+\beta J_3)}}{\sqrt{2G_{1-}(\beta J)\sinh{2\beta J_1}\sinh{2\beta J_3}}} \\
	A_{2341}(K)&=& \frac{\sinh{2(\beta J_2+\beta J_4)}}{\sqrt{2G_{2-}(\beta J)\sinh{2\beta J_2}\sinh{2\beta J_4}}},
\end{eqnarray*}
where
\begin{eqnarray*}
	G_{1-}(\beta J)&=& \cosh{2(\beta J_1+\beta J_3)}+\cosh{2(\beta J_2-\beta J_4)}, \\
	G_{2-}(\beta J)&=& \cosh{2(\beta J_2+\beta J_4)}+\cosh{2(\beta J_1-\beta J_3)}
\end{eqnarray*}
This change also stops the disagreement at low temperatures in these systems. Applying both these changes and the conditions for the $\sigma$-sublattice gives agreement with the simulation results.

Further work is required to understand the implications of the conditions suggested above on the work of Wu and Lin \cite{Wu1989}. In this paper they use a similar approach to the one presented here for the checker-board lattice, so an investigation into this lattice would be useful to see if similar results are obtained. Also as many papers, such as \cite{Strecka2006,Strecka2006a}, extend the results presented here. These should also be studied for similar inconsistencies. Finally, we observed a disagreement between the theoretical predictions and simulation results at low temperatures in our analysis of the system presented in \Fref{fig:NSFunky}. This disagreement remains after the application of the above conditions. As such, further investigation into these systems is required.

\bibliography{bibliography}

\end{document}